\documentclass{ws-ijmpd}
\usepackage[super,compress]{cite}

\usepackage{float}

\usepackage{amsmath,amssymb}

\floatstyle{plaintop}
\restylefloat{table}

\usepackage{color}
\usepackage{ulem}

\begin{document}

\markboth{Satoshi Tsuchida and Masaki Mori}
%{Instructions for Typing Manuscripts (Paper's Title)}
{The gamma--ray spectral feature from Kaluza--Klein dark matter annihilation and its observability}

\catchline{}{}{}{}{}
%
%%%%%%%%%%%%%%%%%%%%%%%%%%%%%%%%%%%%%%%%%%%%%%%%%%%%%%%%%%%%%%%%%%%%

\title{The gamma--ray spectral feature from Kaluza--Klein dark matter annihilation and its observability}

\author{Satoshi Tsuchida$^{*}$}

\address{Department of Physics, Osaka City University, 3--3--138 Sugimoto, Sumiyoshi--ku, Osaka City, Osaka, 558--8585, Japan \\
stsuchida88@gmail.com}

\author{Masaki Mori}

\address{Department of Physical Sciences, Ritsumeikan University, Kusatsu 525--8577, Shiga, Japan}

\maketitle

\begin{history}
\received{Day Month Year}
\revised{Day Month Year}
\end{history}

\begin{abstract}%

The theory of universal extra dimensions involves Kaluza--Klein (KK) particles.
The lightest KK particle (LKP) is one of the good candidates for cold dark matter.
Annihilation of LKP dark matter in the Galactic halo produces high-energy gamma-rays.
The gamma-ray spectrum shows a characteristic peak structure around the LKP mass.
This paper investigates the observability of this peak structure by near-future detectors
taking account of their energy resolution,
and calculates the expected energy spectrum of the gamma-ray signal.
Then, by using the HESS data,
we set some constraints on the boost factor,
which is a product of the annihilation cross section relative to the thermal one and
an uncertain factor dependent on the substructure of the LKP distribution in the Galactic halo, for each LKP mass.
The resulting upper limit on the boost factor is in the range from 1 to 30.
The constraints can be regarded as comparable with
the results of previous work for gamma--ray and electron--positron observation.
However, the observational data for the TeV or higher energy region are still limited,
and the possible LKP signal is not conclusive.
Thus, we expect near-future missions with better sensitivity will clarify
whether the LKP dark matter should exist or not.

\end{abstract}


\ccode{95.35.+d; 95.85.Pw; 98.70.Rz}

\section{\label{sec:intro}Introduction}
\label{intro}

Most of the matter in the Universe is believed to be dark.
The existence of non-luminous matter, so-called dark matter,
was suggested by F. Zwicky in 1930s \cite{Zwicky1933}.
The dark matter problem is one of the most important mysteries in cosmology and particle physics \cite{Bertone2005}.
Various observational data show some indirect evidences supporting the existence of dark matter.
One of the methods to consider what makes dark matter is to suppose that particles predicted by new theories,
which have not yet been detected, are candidates for dark matter \cite{Bergstrom2009}.
One feasible candidate of cold dark matter (CDM) is the weakly interacting massive particles (WIMPs).
By the Planck observational data, the density of CDM is given by
${\Omega}_{\rm{CDM}} h^{2} = 0.1187 \pm 0.0017$ \cite{Ade2014},
where ${\Omega}_{\rm{CDM}}$ is the CDM density parameter of the Universe expressed as a fraction
of the critical density for a flat universe,
and $h$ is the Hubble constant in units of $100 \ {\rm{km \ s^{-1} \ Mpc^{-1}}}$.
The relic density of CDM in the present Universe is determined by the annihilation cross section,
which would be roughly that of the weak interaction if CDM is made of WIMPs.

The theory of universal extra dimensions (UED) is a well--created theory to explain the CDM,
and a popular theory beyond the standard model \cite{Cheng2002},
where {\it{universal}} means that all fields of the standard model can propagate into extra dimensions.
New particles predicted by this theory are called Kaluza-Klein (KK) particles.
Here, we consider the theory of UED containing only one extra dimension,
which is compactified with radius $R$.
At tree level, the KK particle mass is given by \cite{Bergstrom2005}
\begin{eqnarray}
  \label{kaluzakleinmass}
  m^{(n)} = \sqrt{ \left( \frac{n}{R}\right)^{2} + m_{\rm{EW}}^{2}}
\end{eqnarray}
where $n$ is a mode of the KK tower, and $m_{\rm{EW}}$ is a zero mode mass of an electroweak particle.

We assume that the lightest KK particle (LKP) is
a feasible candidate for dark matter, and we denote it $B^{(1)}$,
which is the first KK mode of the hypercharge gauge boson.
Because dark matter should be electrically neutral and stable,
the LKP either does not interact with the standard model particles
or only weakly interacts with them.
In addition, LKP should have a very small decay rate to survive for a cosmological time.
This hypothesis corresponds to the LKP mass $m_{B^{(1)}}$ being in the range
$0.5 \ {\rm{TeV}} \lesssim m_{B^{(1)}} \lesssim 1 \ {\rm{TeV}}$
using the above value for CDM density \cite{Bergstrom2005}.

\begin{figure}[h]
  \begin{center}
    \includegraphics[width=\columnwidth]{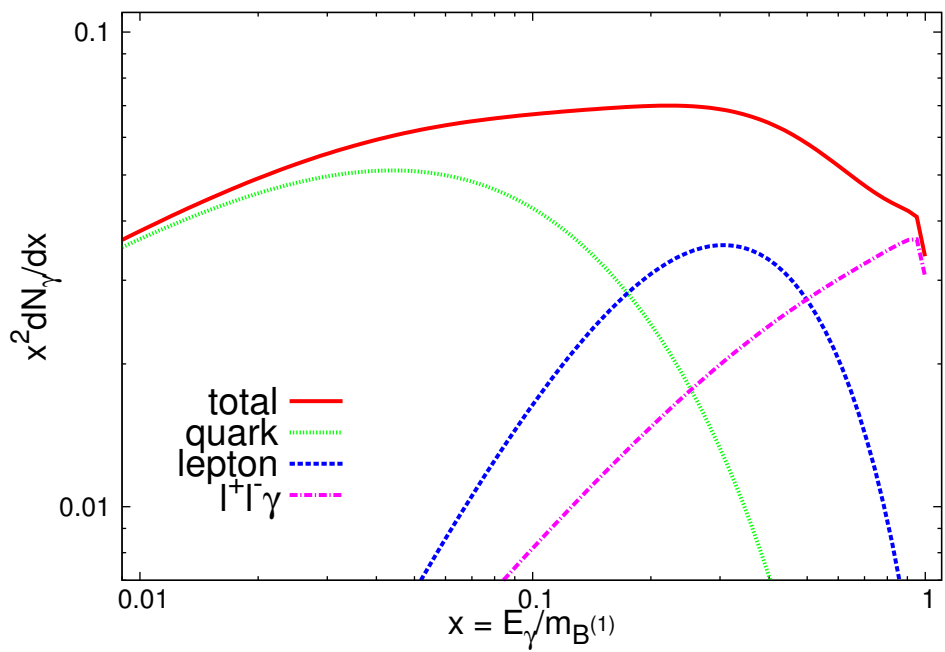}
    \caption{(Color online)
      Gamma-ray spectra of the continuum emission.
      The lines show the number of photons multiplied by $x^{2} = (E_{\gamma} / m_{B^{(1)}})^{2}$ as follows:
      the solid line shows the total number of photons per $B^{(1)} B^{(1)}$ annihilation,
      the dotted line shows the number via quark fragmentation,
      the dashed line shows the number via lepton fragmentation,
      and the dot-dashed line shows the number from the $ l^{+} l^{-} {\gamma}$ component.
      We have assumed $m_{B^{(1)}} = 800$ GeV and mass splitting is 5{\%} at the first KK level.}
    \label{fig:continuum2}
  \end{center}
\end{figure}

There are some LKP annihilation modes which contain gamma-rays as final products.
These include gamma-ray ``lines'' from two-body decays,
and ``continuum'' emission from decay or fragmentation of secondaries.
The cross section for $B^{(1)}$ pair annihilation has been calculated \cite{Bergstromjcap2005},
and we assume the mass splitting is 5{\%} at the first KK level.
In addition, branching ratios into these modes can be calculated for
$B^{(1)}$ pair annihilation \cite{Cheng2002,Bergstrom2005,Servant2003}
and are not dependent on parameters other than $m_{B^{(1)}}$.
The branching ratios are given as follows:
20{\%} for each charged lepton, 11{\%} for each up-type quark,
0.7{\%} for each down-type quark, 1{\%} for each charged gauge boson,
and 0.5{\%} for each neutral gauge boson~\cite{Servant2003, Hooper2003}.
We follow assumptions made in Ref.~[9].
This paper considers three patterns for the continuum:
$B^{(1)}$ pairs annihilate into
(i) quark pairs,
(ii) charged lepton pairs which cascade or produce gamma-rays, and
(iii) two leptons and one photon ($l^{+}l^{-}{\gamma}$).
The gamma-ray spectra of the continuum component for $m_{B^{(1)}} = 800 $~GeV
are reproduced in Fig.~\ref{fig:continuum2}
as per Bergstr{\"o}m $et~al$. \cite{Bergstrom2005}.
In this figure, the solid line shows the total number of photons per $B^{(1)}$ pair annihilations,
the dotted, dashed and dot-dashed lines show the number of photons
via quark fragmentation,
via lepton fragmentation,
and from the $ l^{+} l^{-} {\gamma}$ mode, respectively,
as a function of $ x = E_{\gamma} / m_{B^{(1)}} $.

When $B^{(1)}$ pairs directly annihilate into photon pairs,
they appear as a ``line'' at the $m_{B^{(1)}}$ in the gamma-ray spectrum.
The production cross section of this mode is approximately $ 130 \times 10^{-6} $~pb
for the 5{\%} mass splitting~\cite{Bergstromjcap2005}.
We note that the cross section depends on the mass splitting only weakly.
For example, it is $ 170 \times 10^{-6} $~pb ($ 110 \times 10^{-6} $~pb)
by assuming 1{\%} (10{\%}) mass splitting.
Thus we fix the cross section to be $ 130 \times 10^{-6} $~pb throughout in this paper.
This ``line'' structure is the most prominent signal of KK dark matter,
while in some theories line models are loop-suppressed
and thus usually subdominant (see, e.g. Bringmann $et~al$. \cite{Bringmann2012a}).
This study focuses on the detectability of this ``line'' structure
by near-future detector taking account of their finite energy resolution.

The distribution of dark matter is expected to be non-uniform in the Universe,
and to be concentrated in massive astronomical bodies due to gravity.
The gamma-ray flux from annihilation of dark matter particles
in the Galactic halo can be written as~\cite{Cesarini2004}
\begin{eqnarray}
  \label{gammarayflux}
  {\Phi}_{\gamma} (E_{\gamma}, \psi ) = \frac{ \langle {\sigma} v \rangle }{ 8 \pi M^{2} } \sum_{i} B_{i} \frac{ d N_{\gamma}^{i} }{ d E_{\gamma} } \int_{\rm{line-of-sight}} {\rho}^{2}(l) dl(\psi)
\end{eqnarray}
where $ M $ is the dark matter mass, $ B_{i} $ is the branching ratio into the tree--level annihilation final state $i$,
the function ${\rho}(l)$ is the dark matter density along the line-of-sight $l({\psi})$,
where ${\psi}$ is the angle with respect to the Galactic center,
$dN_{\gamma}^{i} / dE_{\gamma}$ is the gamma--ray spectrum generated per annihilation,
and $\langle {\sigma} v \rangle$ is the total averaged thermal cross section multiplied by the relative velocity of particles.

Now, 
we define a ``boost factor'', $ B_{f} $,
which describes the signal enhancement from dark matter annihilation in the Galactic halo~\cite{Prada2004}.
$ N $--body simulation study given by Navarro--Frenk--White (NFW)~\cite{Navarro1996},
for example, indicates a large $ B_{f} $.
The boost factor $ B_{f} $ is defined as following expression
\begin{eqnarray}
  \label{boostfactor}
  B_f &=& B_{\rho} \times B_{{\sigma}v} \nonumber \\
  &=&  \left( \frac{ \langle {\rho}^{2} (l) \rangle_{{\Delta}V}}{\langle {\rho}_{0}^{2} \rangle_{{\Delta}V}} \right)
\left( \frac{\langle {\sigma}v \rangle }{ 3 \times 10^{-26}~\rm{cm^{3} \ s^{-1}} }\right)_{{\Delta}V}
\end{eqnarray}
where
$ 3 \times 10^{-26}~\rm{cm^{3} \ s^{-1} }$ is the typical cross section multiplied
by velocity expected for thermal production of CDM~\cite{Bergstrom2009},
the volume ${\Delta}V$ is a diffusion scale,
and $ {\rho}_{0} = 0.3 $~GeV cm$^{-3}$ is a typical CDM density around the solar system.
We note that $ B_{\rho} $ corresponds to the following dimensionless integral~\cite{Bergstrom1998}:
\begin{eqnarray}
  \label{eq:jintegral}
  J ( \psi ) = \frac{ 1 }{ 8.5~{\rm{kpc}} } \times \int_{ {\rm{l.o.s}} }
  \frac{ {\rho}^{2} (l) }{ {\rho}_{0}^{2} } dl (\psi)
\end{eqnarray}
Even if CDM density in the halo depends on $ \rho (l) $,
we only concern the ratio of $ \rho (l) $ to $ \rho_{0} $,
so we do not have to take account of details of the halo density profile.
$B_{\rho}$ could be as high as 1000 when taking account of the expected effects of adiabatic compression~\cite{Prada2004}.

Some constraints from observations on the KK dark matter models have been reported.
The Large Area Telescope on board the Fermi Gamma-Ray Space Telescope (Fermi-LAT)
team searched for gamma-ray emission from dwarf spheroidal galaxies
around the Milky Way galaxy and set constraints on dark matter models with non-detection results \cite{Abdo2010}.
The High Energy Stereoscopic System (HESS) array of imaging atmospheric Cherenkov telescopes observed the Sagittarius
dwarf spheroidal galaxy in the sub-TeV energy region and derived a lower limit on the $m_{B^{(1)}}$ of 500~GeV \cite{Aharonian2008}.
These results put constraints on $ B_{\rm{tot}} $ of dark matter halo KK particles.
The present limits allow the maximum value of boost factors
in the range of 2--60 ($ m_{B^{(1)}} = 200 $ GeV) to 600--$ 1.5 \times 10^{4} $ (1000 GeV)
by Fermi-LAT \cite{Abdo2010},
0.8--30 (400 GeV) to 5--160 (1000 GeV) by HESS \cite{Aharonian2008}.

Recent progress in gamma-ray observation has revealed new findings in the Galactic center region
(here we concentrate discussion in the energy region above 10 GeV in our interest).
A point-like high-energy gamma-ray sources at the Galactic center have been observed by HESS \cite{Aharonian2009a}
and other ground-based experiments at energies greater than 100~GeV.
Some authors argued the observed HESS signal might result from annihilation of heavy ($>10$~TeV) dark matter \cite{Cembranos2012},
but since its spectrum is represented well by a power-law plus an exponential cut-off,
it is discussed that the bulk of the emission must have non-dark matter origin \cite{Aharonian2006}.
The Fermi-LAT identified a GeV gamma-ray source 2FGL J1745.6-2858 \cite{Nolan2012} consistent with Sgr $ {\rm{A}^{*}} $,
in arcminute-scale at least above 10 GeV, in accordance with the HESS source \cite{Chernyakova2011}.
The energy spectrum can be explained by a hadronic interaction of relativistic protons
injected to a central source in a power-law plus exponential cutoff spectrum \cite{Chernyakova2011}.
(Note, however, the energy resolution of ground-based atmospheric Cherenkov detectors
like HESS is 15--20\% and narrow features, if any, will be washed out: see below).
Also, an enhancement of around 130 GeV in the energy spectrum of gamma-rays from the
Galactic center region has been reported using the Fermi-LAT data,
which may indicate a possible dark matter annihilation signal
\cite{Bringmann2012b,Finkbeiner2013,Su2012,Weniger2012}.
However, the analysis by the Fermi-LAT collaboration did not confirm
the significance of the line detection and the dark matter 
interpretation is disfavored \cite{Ackermann2012a,Ackermann2013}.
Thus the possible dark matter signal in this energy region is far from
conclusive and clearly we need more sensitive observations.

Near--future missions equipped with high--energy--resolution gamma--ray detectors may resolve
this issue as we discuss in the following.
They could be space--based calorimetric detectors like
the Calorimetric Electron Telescope (CALET)~\cite{Torii2011, Mori2013},
Wukong (DAMPE)~\cite{Chang2014},
HERD~\cite{Huang2016}
or GAMMA--400~\cite{Topchiev2015}.
Former two detectors have been in orbit since 2015 August and 2015 December, respectively,
and the last one is planned to be launched in the middle of 2020s.
These have a percent--level energy resolution and a detection area of one to several thousand square centimeters.
The smallness of the area, compared with atmospheric Cherenkov telescopes,
can be compensated by a longer exposure time in orbit.

In this paper, we analyze the gamma-ray spectral features from $B^{(1)}$ pair annihilation
taking account of the finite energy resolution of gamma-ray detector
and purposefully discuss the observability of the ``line'' at the $m_{B^{(1)}}$.
We then give possible constraints on the boost factor by near-future detector.

\section{The effect of energy resolution}

The gamma-ray flux ${d{\Phi}_{\gamma} ({\Delta}{\Omega})}/{dE_{\gamma}}$ reaching a detector can be expressed as~\cite{Bergstrom2005}
\begin{eqnarray}
  \label{relationxande}
  E_{\gamma}^{2} \frac{d{\Phi}_{\gamma}({\Delta}{\Omega})}{dE_{\gamma}} = K \times B_{f} \times x^{2} \frac{dN_{\gamma}}{dx},
\end{eqnarray}
where ${\Delta} {\Omega}$ is the angular acceptance of the detector,
\begin{eqnarray}
  \label{constfactor}
  K \simeq 3.5 \times 10^{-8}~{\rm{m^{-2} s^{-1} TeV}} \left( \frac{0.8 {\rm{TeV}}}{m_{B^{(1)}}} \right) \nonumber \\
  \times \left( \frac{ \langle \sigma v \rangle_{\rm{LKP}}}{3 \times 10^{-26}~\rm{cm^{3} \ s^{-1} }} \right)
  \langle J_{\rm{GC}} {\rangle}_{{\Delta}{\Omega}} {\Delta}{\Omega},
\end{eqnarray}
where $\langle \sigma v \rangle_{\rm{LKP}} \simeq 3 \times 10^{-26} (0.8~{\rm{TeV}} / m_{B^{(1)}})^{2}~\rm{cm^{3} \ s^{-1} }$ is the cross section multiplied by velocity expected for thermal production of LKP~\cite{Bergstrom2005},
and $\langle J_{\rm{GC}} {\rangle}_{{\Delta}{\Omega}}$ is a dimensionless line-of-sight integral averaged over ${\Delta}{\Omega}$, which corresponds to Eq.~(\ref{eq:jintegral}).
If we assume an NFW profile,
$\langle J_{\rm{GC}} {\rangle}_{{\Delta}{\Omega}} {\Delta}{\Omega}$ equals to 0.39 for a ${\Delta}{\Omega}= 10^{-4}$~\cite{Cesarini2004},
where the $ {\Delta} {\Omega} $ is assumed to be reasonable value both for
the angular resolution of CALET (0.2 - $ 0.3^{\circ} $)~\cite{Mori2013}
and the observed localization of the Galactic center source observed by Fermi-LAT~\cite{Chernyakova2011}.
In this case $dN_{\gamma}/dx$ includes both the continuum and line components.

\begin{figure}[t]
  \begin{center}
    \includegraphics[width=120mm]{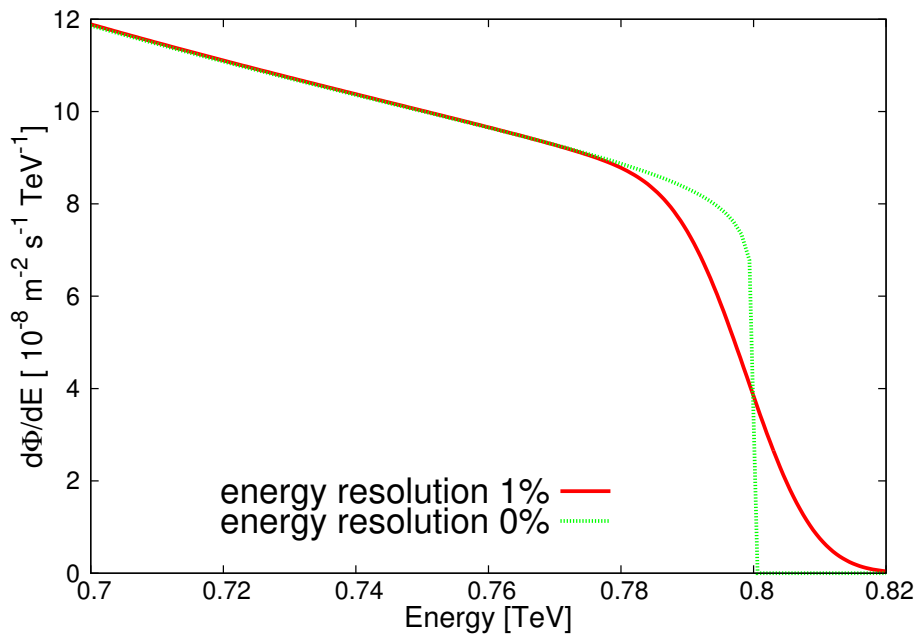}
    \caption{(Color Online)
      Gamma-ray spectrum of the continuum taking account of the energy resolution assuming $m_{B^{(1)}} = 800$ GeV.
      The solid line assumes an energy resolution of 1{\%} with a Gaussian distribution,
      and the dotted line does not include the effect of energy resolution,
      as per the solid line in Fig.~\ref{fig:continuum2}.
      The assumed boost factor is 100.}
    \label{fig:continuumenereso2}
  \end{center}
\end{figure}

Now, we discuss the effect of energy resolution of detectors.
If the measured energy dispersion for mono-energetic gamma-rays 
behaves as a Gaussian distribution and the energy resolution of the detector is finite,
the measured gamma-ray spectrum is blurred.
This effect is shown in Fig.~\ref{fig:continuumenereso2} for the ``continuum'' component
assuming the 1{\%} energy resolution.
Here we draw the curve assuming the following equation
\begin{eqnarray}
  \label{energygaussian}
  g(E) \propto \int f(E') \times {\rm{exp}} \left[ - \frac{(E-E')^{2}}{2{\sigma}_{E}^{2}} \right] dE' ,
\end{eqnarray}
where $f(E')$ corresponds to a function shown by the solid line in Fig.~\ref{fig:continuum2},
and ${\sigma}_{E}$ is the energy resolution.

Next we analyze how the ``line'' from the $B^{(1)}$ pair annihilation into photon pairs looks above the ``continuum''.
\begin{figure}[t]
  \begin{center}
    \includegraphics[width=120mm]{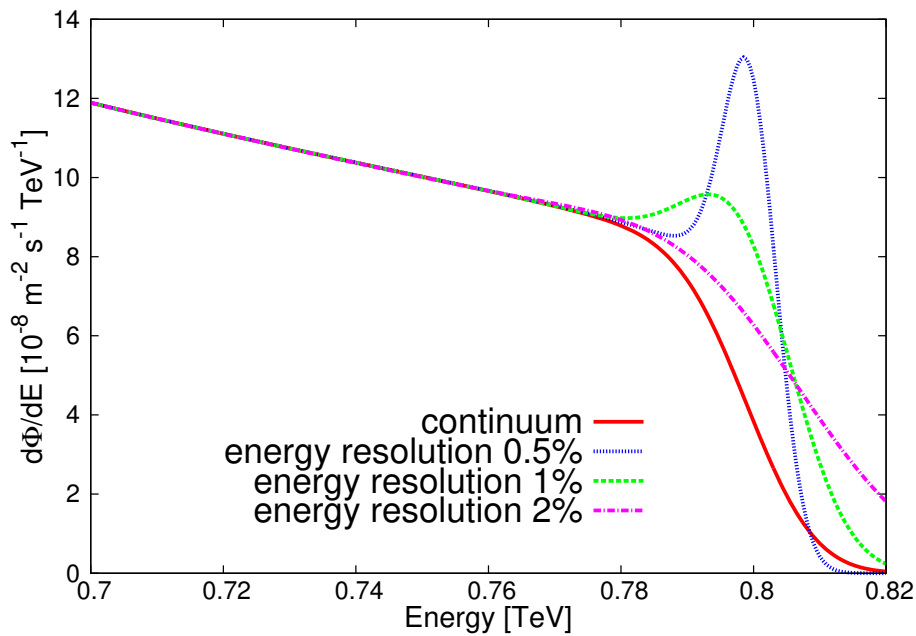}
    \caption{(Color Online)
      Gamma-ray spectra of continuum plus line diffused by the energy resolution assuming $m_{B^{(1)}} = 800$ GeV.
      The solid line shows the continuum component only, assuming the energy resolution of 1{\%},
      while the dotted, dashed and dot-dashed lines show the continuum plus line components assuming
      energy resolution values of 0.5{\%}, 1{\%} and 2{\%} respectively.
      The assumed boost factor is 100.}
    \label{fig:con1perwab}
  \end{center}
\end{figure}
In Fig.~\ref{fig:con1perwab}, the solid line shows the continuum component only with an energy resolution of 1{\%},
and the patterned lines show ``line'' plus ``continuum'' spectra for different energy resolutions:
the dotted line, dashed line and dot-dashed line show the spectra
when the energy resolution is 0.5{\%}, 1{\%} and 2{\%} with the Gaussian distribution respectively,
assuming the boost factor $B_{f} = 100$.
We also point out that the peak energies of the expected spectra ($ E_{\rm{peak}} $) are
0.2{\%} and 0.9{\%} smaller than $ m_{B^{(1)}} $,
for 0.5{\%} and 1{\%} energy resolution, respectively.
For 2{\%} energy resolution, the peak structure is difficult to see.

\begin{figure}[t]
  \begin{center}
    \includegraphics[width=120mm]{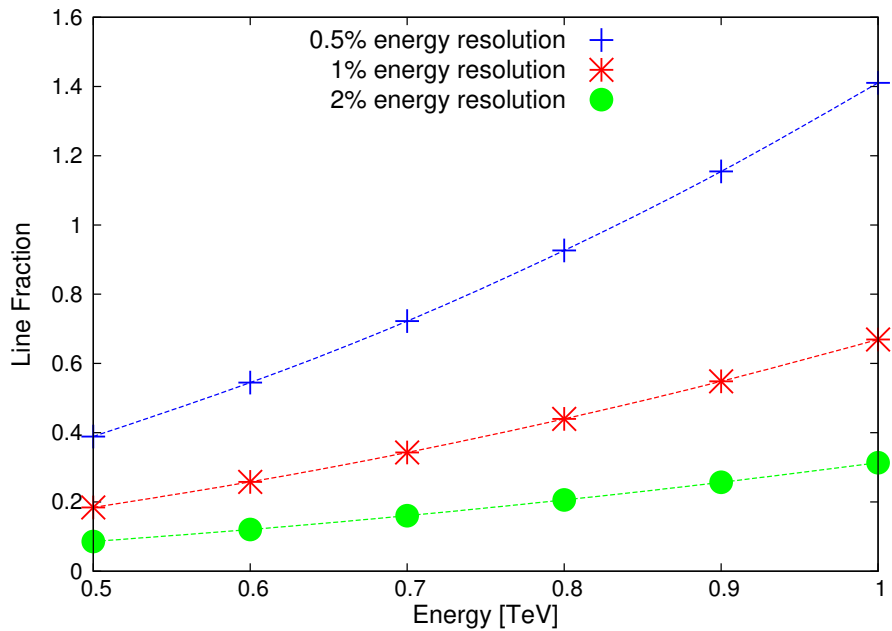}
    \caption{(Color Online)
      The line fraction as a functions of $m_{B^{(1)}}$,
      assuming an energy resolution of 0.5{\%}, 1{\%} and 2{\%}.
      The dashed curve is drawn to guide to the eyes.}
    \label{fig:linefraction}
  \end{center}
\end{figure}

To investigate the tendency of the line component quantitatively,
we consider the line to continuum ratio, which is referred to as ``Line fraction ($LF$)''.
$LF$ is defined as
\begin{eqnarray}
  \label{eq:eqwidth}
  LF = \frac{ \sum_{i} F_{i}^{l} }{ \sum_{i} F_{i}^{c} }
\end{eqnarray}
where $F_{i}^{c}$, $F_{i}^{l}$ are the fluxes of the continuum component
and the line component of the $i$-th energy bin, respectively,
and the energy bin width is set to 0.5 GeV.
The summation runs from the lower to the upper energy limit of the observed line.
This range is taken as $ - 3 {\sigma}_{E} $ to $ + 3 {\sigma}_{E} $ for each $m_{B^{(1)}}$,
since the flux above $m_{B^{(1)}}$ drops rapidly.
The result is shown in Fig.~\ref{fig:linefraction} as a function of $m_{B^{(1)}}$.
In this figure, we can see that the value of LF increases as $m_{B^{(1)}}$ becomes heavier,
which implies characteristic peak structure is clearer for heavier $m_{B^{(1)}}$.
Our definition of $LF$ can be regarded as a similar quantity to the
`equivalent width' used in line spectroscopy,
so that one may require $ LF > 1 $ as a criterion to detect a line structure definitively.
This corresponds to the current case of 0.5{\%} or better energy resolution
and $ m_{B^{(1)}} $ of 800~GeV or heavier.

\begin{figure}[t]
  \begin{center}
    \includegraphics[width=120mm]{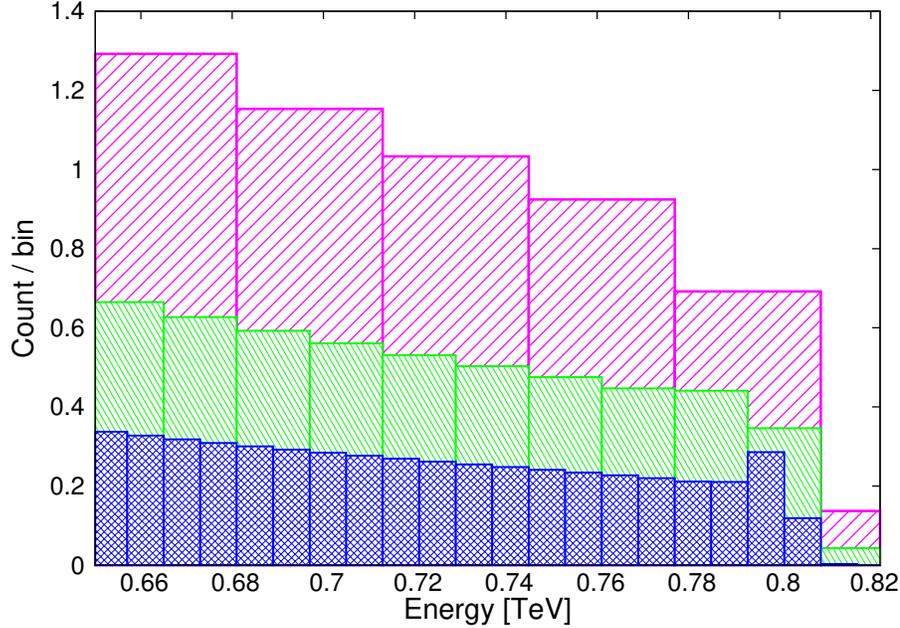}
    \caption{(Color Online)
      Expected count spectra near the peak assuming energy resolutions of 0.5{\%}, 1{\%} and 2{\%} assuming $m_{B^{(1)}} = 800$ GeV
      and $ 3 \times 10^{8}~{\rm{m^{2}~s}}$ exposure.
      The bin widths of histograms are twice as much as 0.5{\%}, 1{\%} and 2{\%} of the $m_{B^{(1)}}$, respectively.
      The assumed boost factor is 100.}
    \label{fig:countbins}
  \end{center}
\end{figure}

We can transform the spectra into counts to be observed by gamma-ray detectors.
This is accomplished through multiplying by a factor of $3 \times 10^{8}$~m$^{2}$~s
for an assumed observation time of 10 yr = $ 3 \times 10^{8}$~s and an effective area of 1 ${\rm{m}^{2}}$.
These values are becoming realistic with the characterization for space--based gamma--ray detectors,
such as HERD.
When analyzing observational data, the energy bin width must be specified.
The bin widths of histograms are set to twice as much as 0.5{\%}, 1{\%} and 2{\%} of $m_{B^{(1)}}$
in order to match the each energy resolution.
The resulting histograms for $ m_{B^{(1)}} = 800$~GeV are shown in Fig.~\ref{fig:countbins},
where plots of the three cases corresponding to energy resolutions of 0.5{\%}, 1{\%} and 2{\%} are shown.
The figure shows that if the energy resolution of the detector
becomes 2{\%} or worse, the characteristic peak indicating the $m_{B^{(1)}}$ will be diffused,
and making it hard to resolve into the line and continuum components.
Here, we do not consider for the systematic error, because the shape of spectral features will not be changed.

\begin{figure}[t]
  \begin{center}
    \includegraphics[width=120mm]{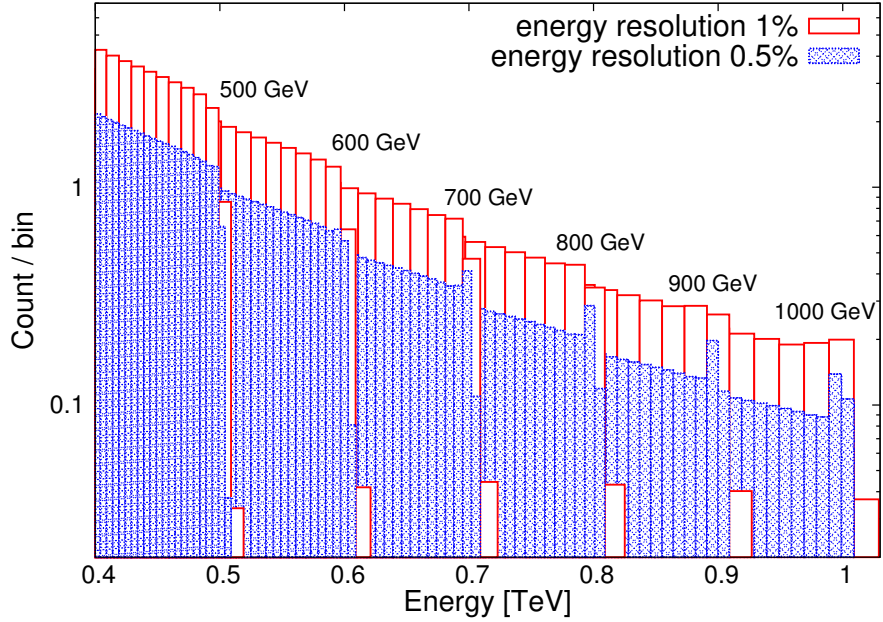}
    \caption{(Color Online)
      Expected count spectra, assuming energy resolutions of 0.5{\%} and 1{\%}, and $ 3 \times 10^{8}~{\rm{m^{2}~s}}$ exposure.
      The data spaces are twice as much as 0.5{\%} and 1{\%} of the $m_{B^{(1)}}$.
      The assumed boost factor is 100.}
    \label{fig:countbins_compare}
  \end{center}
\end{figure}

Now, we vary the mass from 500 GeV to 1000 GeV in 100 GeV intervals,
and calculate the count spectrum for each mass.
The results are shown in Fig.~\ref{fig:countbins_compare},
which shows that the characteristic peak structure is visually clearer when $m_{B^{(1)}}$ is heavier.
That is, the line component becomes relatively larger since the continuum component decreases for heavier $m_{B^{(1)}}$.

\section{Discussion}
\label{sec:constraint_gamma}

We now discuss the observability of the LKP signal in near-future detectors,
taking account of present gamma--ray observations.
That is, we give estimates for the accessible range of the boost factor
based on upper limits on extra component in energy spectra.
Here, we consider the gamma--ray spectrum of HESS J1745-290 located near the center of the Galaxy.
This gamma--ray source can be identified as the Galactic center, Sgr A$^*$~\cite{Chernyakova2011}.
Its energy spectrum above 200~GeV is given by~\cite{Abramowski2016}
\begin{eqnarray}
  \label{eq:background}
  \frac{d{\Phi}}{dE} = \left( 2.55 \pm 0.04_{\rm{stat.}} \pm 0.37_{\rm{syst.}} \right) \left( \frac{ E }{ \rm{TeV} } \right)^{-2.14 \pm 0.02_{\rm{stat.}} \pm 0.10_{\rm{syst.}} } \  \nonumber \\
         \times \ {\rm{exp}} \left[ - \frac{E}{ \left( 10.7 \pm 2.0_{\rm{stat.}} \pm 2.1_{\rm{syst.}} \right)~{\rm{TeV}}} \right] \ \ \ \ \ \  \nonumber \\
\times 10^{-8} \ \ \ {\rm{TeV}^{-1}} \ {\rm{m}^{-2}} \ {\rm{s}^{-1}}.\ \ \ \ 
\end{eqnarray}
Note that with the energy resolution of HESS (15--20{\%}),
which is a system of atmospheric Cherenkov telescopes on the ground,
the LKP ``line'' signal is broadened and hard to detect,
but the ``continuum'' signal could produce some structure in the energy spectrum as below.

Now, we investigate the upper limit on the boost factor, $B_{f}$, with the HESS J1745-290 spectrum.
To do this, we assume the spectrum is composed of a background spectrum,
represented by a power--law with an exponential cutoff, and a LKP signal.
Then we employ the least--squares method to set limits on $ B_{f} $.
Thus the model is given by
\begin{eqnarray}
  \label{eq:model}
  \frac{ d{\Phi}_{\gamma} }{ d E_{\gamma} } &=& \frac{ d{\Phi}_{\gamma}^{{\rm{Bkgd}}} }{ d E_{\gamma} } + B_{f} \frac{ d{\Phi}_{\gamma}^{{\rm{LKP}}} }{ d E_{\gamma} }
\end{eqnarray}
with a background gamma--ray spectrum, $ d {\Phi}^{\rm{Bkgd}}_{\gamma} / d E_{\gamma} $:
\begin{eqnarray}
  \frac{ d{\Phi}_{\gamma}^{{\rm{Bkgd}}} }{ d E_{\gamma} } &=& C_{B} \left( \frac{ E }{ {\rm{TeV}} } \right)^{\Gamma_{B}} \exp \left[ - \frac{ E }{ E_{B}~{\rm{TeV}} } \right] \times 10^{-8}~{\rm{m^{-2}~s^{-1}~{TeV}^{-1} } }
  \label{eq:backgroundflux}
\end{eqnarray}
where $C_{B}$, ${\Gamma}_{B}$, and $ E_{B} $ are a coefficient, a power--law index, and cutoff energy of the background spectrum.
Here we assume that the energy resolution of detector is 20{\%}, which corresponds to that of HESS,
so the ``LKP Flux'', $ d{\Phi}_{\gamma}^{{\rm{LKP}}} / d E_{\gamma} $,
has the line plus continuum components blurred by 20{\%} energy resolution.

The goodness of the model fit to the HESS data can be tested by the sum
\begin{eqnarray}
  {\chi}^{2} = \sum_{i} \frac{ ( {\rm{data}} - {\rm{model}} )^{2}_{i} }{ {\sigma}_{i}^{2} }
  \label{eq:chisq_hess}
\end{eqnarray}
where ``data'' is the HESS data point, ${\sigma}_{i}$ is its error,
and ``model'' is given by Eq.~(\ref{eq:model}).
The index $ i $ runs the data points in the energy range between about 200~GeV to 20~TeV.
A number of degrees of freedom is 21
(= a number of data points (25) minus unknown parameters ($ 4 = C_{B} $, $ {\Gamma}_{B} $, $ E_{B} $ and $ B_{f} $)).
Thus, ${\chi}^{2} < 38.9$ is required for the fit to be consistent with the HESS data at 99{\%} confidence level.
We calculate $ {\chi}^{2} $ values for various parameter sets, where
we vary the coefficient $ C_{B} $ from 2.05 to 2.55 in 0.01 step,
the index $ {\Gamma}_{B} $ from -2.25 to -1.75 in 0.01 step,
the cutoff energy $ E_{B} $ from 7.9 to 13.5 in 1.4 step,
and the boost factor $ B_{f} $ from 0 to 100 in 1 step.
The results for this calculation are shown in Table~\ref{table:hessparameter},
which indicates the maximally allowed $B_{f}$ for each LKP mass, $ m_{B^{(1)}} $, at 99{\%} confidence level ($ {\chi}^{2} < 38.9 $).
The combinations of parameter sets $ ( C_{B}, {\Gamma}_{B}, E_{B} ) $,
which satisfy the condition $ {\chi}^{2} < 38.9 $,
for each $ m_{B^{(1)}} $ are not unique,
but we show the maximum values of $ B_{f} $,
which is our main concern, in Table~\ref{table:hessparameter}.
\begin{table}[t]
  \begin{center}
    \caption{The maximally allowed $B_{f}$ for each LKP mass, $ m_{B^{(1)}} $, at 99{\%} confidence level ($ {\chi}^{2} < 38.9 $).}
    \label{table:hessparameter}
    \begin{tabular}{ c c | c c } \hline \hline
     $ m_{B^{(1)}} $~[GeV] & $ B_{f} $ & $ m_{B^{(1)}} $~[GeV] & $ B_{f} $ \\ \hline
              500          &     1     &          1100         &      12   \\
              600          &     2     &          1200         &      16   \\
              700          &     3     &          1300         &      20   \\
              800          &     5     &          1400         &      24   \\
              900          &     7     &          1500         &      30   \\
	      1000         &     9     &                       &           \\ \hline \hline
    \end{tabular}
  \end{center}
\end{table}

\begin{figure}[t]
  \begin{center}
    \includegraphics[width=120mm]{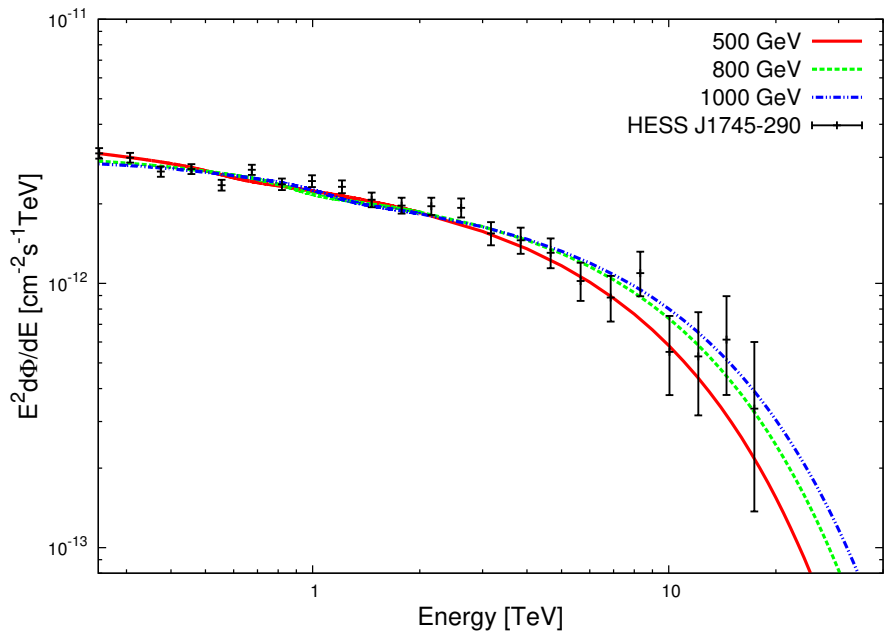}
    \caption{(Color Online)
      Comparison the LKP plus background spectra assuming some parameter sets for ($ C_{B} $, $ {\Gamma}_{B} $, $ B_{f} $)
      and 20{\%} energy resolution with HESS observational data.
      The solid, dashed and dot--dashed lines are the spectra fits to the HESS data at 99{\%} confidence level with
      $ ( C_{B}, {\Gamma}_{B}, E_{B}, B_{f} ) = ( 2.54, -2.09, 7.9, 1 ) $, $ ( 2.37, -2.04, 9.3, 5 ) $, $ ( 2.29, -2.05, 10.7, 9 ) $
      for $ m_{B^{(1)}} = 500 $, 800, 1000~GeV, respectively.
    }
    \label{fig:hesscompare_data}
  \end{center}
\end{figure}

Examples of the resulting LKP plus background gamma--ray spectra assuming
some parameter sets for
($ C_{B} $, $ {\Gamma}_{B} $, $ E_{B} $, $ B_{f} $)
and 20{\%} energy resolution with HESS observational data are given in Fig.~\ref{fig:hesscompare_data}.
The solid, dashed and dot--dashed lines are the model fits compatible with the HESS data at 99{\%} confidence level with
$ ( C_{B}, {\Gamma}_{B}, E_{B}, B_{f} ) = ( 2.54, -2.09, 7.9, 1 ) $, $ ( 2.37, -2.04, 9.3, 5 ) $, $ ( 2.29, -2.05, 10.7, 9 ) $
for $ m_{B^{(1)}} = 500 $, 800, 1000~GeV, respectively.
This figure implies that the LKP plus background spectra decline slightly around assumed LKP mass.
However, the characteristic peak structure around $ m_{B^{(1)}} $ is not seen clearly,
because of blurs caused by 20{\%} energy resolution.

\begin{figure}[t]
  \begin{center}
    \includegraphics[width=\columnwidth]{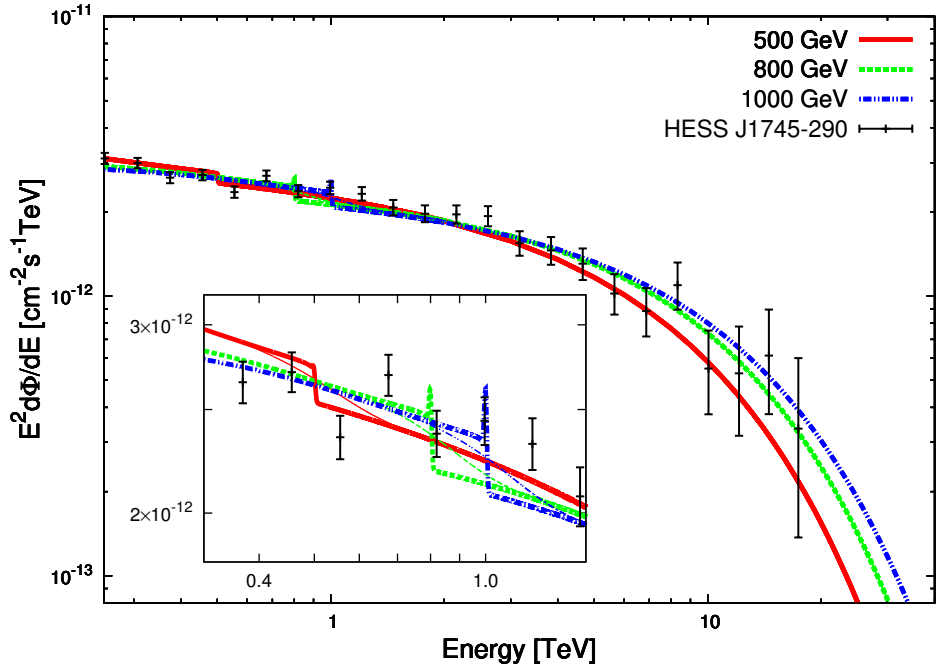}
    \caption{(Color Online)
      Comparison the LKP plus background spectra with HESS data (points).
      The thick solid, dashed and dot--dashed lines show spectra assuming 0.5{\%} energy resolution,
      and the same parameter sets as Fig.~\ref{fig:hesscompare_data}.
      Thin lines are the same as Fig.~\ref{fig:hesscompare_data} for comparison.
      }
    \label{fig:hesscalet}
  \end{center}
\end{figure}

On the other hand, we can think of the case that gamma--rays are detected by 
space--based calorimetric detectors like CALET or Wukong (DAMPE).
With their fine energy resolution, we expect that the peak structure can be detected around $ m_{B^{(1)}} $.
Although space--based detectors have a disadvantage compared with HESS in their small effective area,
we may anticipate that they will observe enough number of gamma--ray events in many years in orbit.
When a large number of gamma--ray events with 0.5{\%} energy resolution are accumulated,
we may see the LKP plus background spectra as shown in Fig.~\ref{fig:hesscalet}.
In this figure, the thick solid, dashed and dot--dashed lines show the spectra to be observed with
0.5{\%} energy resolution for the same parameter sets for $ C_{B} $, $ \Gamma_{B} $, $ E_{B} $ and $ B_{f} $
in Fig.~\ref{fig:hesscompare_data}.
With a large statistics,
the characteristic peak structure will appear clearly around the LKP mass energy region.

It is important to compare our results with those given in previous studies.
The detectability of gamma--rays from dark matter annihilation has been discussed in many literatures.
For instance, Bergstr{\"o}m $et~al$.~\cite{Bergstrom2012} discussed the relation between the mass of dark matter and cross section,
and gave an upper limit for the cross section by using experimental data.
In our calculation, for $m_{B^{(1)}}$ = 1000~GeV and 0.5{\%} energy resolution
the upper limit on $ B_{f} $ would be about $ 10 $.
On the other hand, the result of Ref.~\cite{Bergstrom2012} indicates the upper limit on
the cross section is about $ 9 \times 10^{-28}~{\rm{cm^{3}\ s^{-1}}}$
assuming 1000~GeV dark matter mass for HESS--II observation.
Assuming $ 130 \times 10^{-6} $ pb for the cross section for annihilation into photon pairs,
this upper limit corresponds to about $ B_{f} = 230$.
Thus, with high--energy--resolution and good--statistics gamma--ray observations,
we can set more stringent upper limit on $ B_{f} $ compared with atmospheric Cherenkov telescopes case.

In addition, the relevant value for the boost factor, $ B_{f} $,
to explain the total electron plus positron spectrum
observed by AMS--02 is given in range from 0 to 240 by assuming specific dark matter halo density model,
halo propagation model and $ m_{B^{(1)}} = 1000 $~GeV~\cite{Tsuchida2017}.
This constraint can be comparable with our result for analysis of the gamma--ray spectrum.

In this paper, we point out that space--based calorimetric detectors,
which are smaller in effective area but has better energy resolution compared with those of HESS-II,
has a chance to detect the ``line'' component,
not just setting an upper limit on the boost factor.
Bringmann $et~al$.~\cite{Bringmann2011} discuss the detectability for generic models and treat
the line and continuum component separately, assuming observation by atmospheric
Cherenkov telescopes with 10{\%} energy resolution.
Here, we consider the specified model, which is the annihilation of LKP dark matter
and includes the line and continuum component, assuming high--energy--resolution detectors.

\section{Conclusion}
\label{sec:conclusion}

In this paper, we discussed the observability of characteristic spectral feature
appearing in secondarily produced gamma rays from annihilation of LKP dark matter near the Galactic center.
Here, we conclude the results from our analyses.

Energy resolution plays a key role in detecting the line structure of the gamma-ray spectral features
expected from annihilation of LKP dark matter as predicted by UED theories.
This paper investigated the effects of energy resolution of gamma-ray detector and calculated the expected count spectrum.
The predicted gamma-ray spectrum is the sum of the continuum and the line corresponding to the LKP mass, $m_{B^{(1)}}$,
but this characteristic structure is diluted when we take account of finite energy resolution of detectors
as shown in Fig.~\ref{fig:continuumenereso2} and Fig.~\ref{fig:con1perwab}.
Further, if we assume the exposure
(effective area multiplied by observation time) of near-future detectors,
count statistics will be the final limiting factor as per Fig.~\ref{fig:countbins}.
The characteristic peak indicating $m_{B^{(1)}}$ would be diffused if the energy resolution is 2{\%} or worse.
In addition, if $m_{B^{(1)}}$ is heavy, the observed gamma-ray spectrum will show the characteristic peak clearly
because the continuum component decreases relative to the line component,
as is shown in Fig.~\ref{fig:linefraction} and Fig.~\ref{fig:countbins_compare}.

The obtained upper limits on $B_{f}$ 
based on the analysis of HESS observation are given in Table~\ref{table:hessparameter},
and these values are consistent with previous works
as discussed in Section~\ref{sec:constraint_gamma}.
In addition, if we can detect the LKP signal with 0.5{\%} energy resolution detector,
we would see the characteristic structure around the $m_{B^{(1)}}$, as is shown in Fig.~\ref{fig:hesscalet},
when large number of photons are detected.

If the characteristic structure in gamma-ray flux is observed by new and future missions,
like CALET~\cite{Torii2015}, Wukong~\cite{Chang2014} and GAMMA-400~\cite{Topchiev2015},
we may conclude dark matter is made of LKP.
It would be a conclusive evidence for the existence of extra dimensions.

\section*{Acknowledgments}

We would like to thank Mr. Akihiko Kawamura and Dr. Fumihiro Matsui
for useful discussions and helpful comments.

\end{document}